# Semantic integration and analysis of clinical data

Hong Sun, Kristof Depraetere, Jos De Roo, Boris De Vloed,

Giovanni Mels, Dirk Colaert

Advanced Clinical Applications Research Group, Agfa HealthCare, Gent, Belgium
`{hong.sun, kristof.depraetere, jos.deroo, boris.devloed, giovanni.mels, dirk.colaert }@agfa.com`

**Abstract.** There is a growing need to semantically process and integrate clinical data from different sources for Clinical Data Management and Clinical Decision Support in the healthcare IT industry. In the clinical practice domain, the semantic gap between clinical information systems and domain ontologies is quite often difficult to bridge in one step. In this paper, we report our experience in using a two-step formalization approach to formalize clinical data, i.e. from database schemas to local formalisms and from local formalisms to domain (unifying) formalisms. We use N3 rules to explicitly and formally state the mapping from local ontologies to domain ontologies. The resulting data expressed in domain formalisms can be integrated and analyzed, though originating from very distinct sources. Practices of applying the two-step approach in the infectious disorders and cancer domains are introduced.

## 1    Introduction

A decade ago formal semantics, the study of logic within a linguistic framework, found a new means of expression, i.e. the World Wide Web and with it the Semantic Web [1]. The Semantic Web provides additional capabilities that enable information sharing between different resources which are semantically represented. It consists of a set of standards and technologies that include a simple data model for representing information (RDF) [2], a query language for RDF (SPARQL) [3], a schema language describing RDF vocabularies (RDFS) [4], a few syntaxes to represent RDF (RDF/XML [20], N3 [19]) and a language for describing and sharing ontologies (OWL) [5]. These technologies together build up the foundation to formally describe, query and exchange information with explicit semantics.

Meanwhile, after decades of development of electronic medical records in clinical information systems (CIS), the integration of patient records between different CISs has become a rising request. The development of standard clinical information models is an attempt to tackle the storage and exchange of clinical data. Standards like HL7 [6], openEHR [7], ISO 13606 [8], etc., are developed to keep patient records in struc-tured formats. Although these standards improve interoperability, they still leave room for interpretation because unlike the Semantic Web languages they lack a model theory [4] based on logic and math. In addition, as there are so many electronic health record (EHR) standards developed so far, different CISs might use different standards, which in the end does not contribute to seamless exchanging clinical data.

In order to integrate data represented with different EHR standards, Semantic Web technology is used to create a common ontology to which different EHRs could be mapped. In practice most CISs store clinical data in their own proprietary data structure and build interfaces to export patient data following certain standards. Therefore, we propose to use Semantic Web technology to formalize directly from the database structure of local CISs. It is our belief that the sooner one can work with explicit and formalized data, the better. The use of conventional non-semantic standards to export clinical data towards a data repository (data warehouse) and then raise the data to the formal level introduces a black box with implicit assumptions and software code eliminates the possibility to ever come back to the originating data source.

We started our research on applying Semantic Web technologies in the healthcare environment 10 years ago, driven by the need to enhance clinical decision support (CDS) and clinical workflow. We deem that the prerequisite for CDS is the ability to integrate clinical data from different sources. Such data integration implies that semantics in different sources are translated into a lingua franca of a common ontology.

This paper introduces our experiences in formalizing CISs based on their own data structures. We focus our discussions on formalizing clinical data stored in relational database (RDB) as most CISs store their data in a RDB. The database schema of a CIS is mapped to domain ontologies, which reuse many existing ontologies. In order to scale down the big semantic gap between the local database schema and the domain ontologies, we propose to use a two-step formalization approach: we use local ontologies as an intermediate to bridge the above mentioned semantic gap. These local ontologies are generated based on the database structures of local CISs and are therefore a formal representation thereof. They are further mapped to global ontologies with explicit rules. Such a two-step formalization approach has been implemented in the Debug-IT project [9] for monitoring antimicrobial resistance, and in the HIT4CLL project [10] in the cancer domain of chronic lymphocytic leukemia (CLL).

The rest of this paper is organized as follows: in Section 2, we discuss the related work, and a comparison between a one-step approach and a two-step approach in semantic formalization is given in Section 3. Section 4 introduces the structure of the two-step formalization approach, and implementations are presented in Section 5. Conclusions are given in Section 6.

## 2 Related Work

Much research is carried out in semantically formalizing and connecting different data sources into a common one. The RDB2RDF Working Group under W3C [11] is standardizing a language [18] for expressing customized mappings from RDBs to RDF datasets. There are good tools developed, e.g. D2R [12], which either help to dump RDB data as RDF, or allow SPARQL queries on an RDB via a SPARQL endpoint.

Much research formalizes data directly with ontology formalisms of their target domains [14]. However, due to the semantic gaps between local systems and the domain ontology, it is difficult to directly model the local data using a domain ontology, especially when integrating different data sources together. Bizer et al propose to use a step-by-step integration approach in integrating large scale data [13] [25], so as to decrease the heterogeneity over time, and we share the same view in this paper. In their linked data integration framework [25], Bizer et al first replicate web data locally keeping their original semantics, and then translate them with a target domain ontology [13] in a second step. Their research follows the formalization layers presented in the Semantic Web Stack [24], and are also tested in processing large amount of data. However, their research mainly focuses on building a general framework for integrating web data, to the best of our knowledge, there are no reports on using these tools to formalize real world complex systems, e.g. processing complex clinical data in a running CIS, as what we report in this paper.

There is a strong desire in formalizing clinical data so that EHRs could be formalized and integrated from different clinical information systems [17]. The semantic research in the clinical domain is mostly built on existing EHR standards. In [14], Gonçalves et al build up a domain ontology in ECG and map schemas of three ECG standards to the domain ontology respectively. But it turns out to be difficult to develop a common ontology to which different standards can directly be mapped. In [15], local ontologies for openEHR and ISO 13606 are developed together with a common ontology. However the data transformation is only achieved through syntactic mapping between the archetypes of openEHR and ISO 13606. This is not a semantic mapping because the ontologies of these two standards are not used.

The semantic gaps between different clinical standards prohibit their usage in formalizing and integrating clinical data from different resources, as reflected in [14] [15]. In addition, a clinical information system, e.g. the ORBIS® CIS system used in this paper, normally stores clinical data in its own data structure instead of using existing standards. It is better to directly formalize on the original data in its own data structure, so as to avoid unnecessary interpretations. Holford et al take this approach [16], they formalized a cancer database based on its own data structure, and further integrated it with the RDF dump of a gene ontology. We take a similar approach in this paper starting our formalizations directly on the data stored in CIS database with semantics of the local database. Moreover, we continued our formalization with explicit rules to convert the data expressions using domain ontology. We also use analysis rules to infer more clinical information based on the converted data.

## 3 Semantic Formalization: One Step Versus Two Steps

To make clinical information systems semantically interoperable [17], all relevant meaning from data sources has to be captured in a lingua franca or unifying formalism, realizing integrated data, whereupon various operations can be performed, i.e. 1) asserting the formalized data, 2) formally querying the data to retrieve direct results, and/or 3) executing rules, providing logical implication for inferring new facts from this data, and 4) providing logic proof for the results.

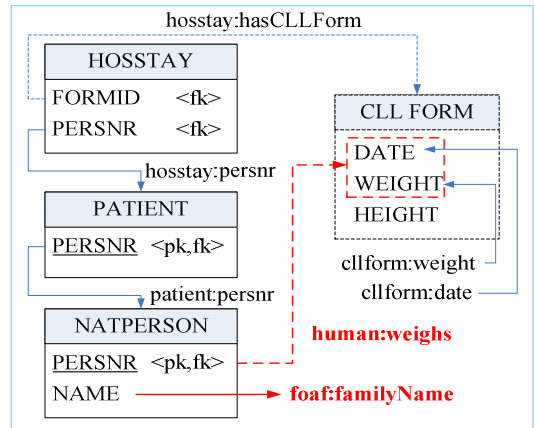

**Fig. 1.** Database schema in a real clinical information system

Creating a unifying formalism in the shape of one or more domain ontologies is one thing. Connecting these ontologies with the source database schemas and values is another. Database schemas usually have very local, specific – and sometimes awkward – semantics due to a variety of reasons, e.g. technical performance, the merging of different products into one, or specific way of processing clinical information (in a clinical database).

Figure 1 depicts a portion of the database schemas of a real CIS together with the formalization of the displayed tables. The solid lines indicate direct links and the dotted lines indicate indirect links. The "Hosstay", "Patient" and "Natperson" are all tables in a RDB, while the "CLL Form" table is a virtual table indirectly connected to the "Hosstay" table. FOAF (http://xmlns.com/foaf/0.1/) and Human (http://eulersharp.sourceforge.net/2003/03swap/human#) are the domain ontologies to be mapped to. Using an one step approach, it is easy to map the 'name' column in the "Natperson" table to foaf:familyName. However, it becomes more complex to state a property human:weighs which needs to map to a blank node consisting of the 'weight' and 'date' columns in the "CLL Form" table, to formalize the relationship between the weight value and date of measurement (explained in more detail below). Therefore we propose to use a two-step approach to formalize CISs in a more expressive way.

Figure 2 compares the one step formalization approach and the two-step formalization approach. Both approaches target to formalize the DB schemas in the operational world to domain formalisms in the formal world. In order to map to domain formalism, two mapping processes have to be completed either in one step or two steps: 1) Mapping the operational world to the formal world, this is the RDB-to-RDF mapping when source data is stored in a RDB. 2) Mapping the database semantic to the domain formalisms. In both the one step and two-step approaches, domain formalisms (ontology A and ontology B2) may utilize existing ontologies to build up domain ontologies, however, their expressivity differs, as explained below.

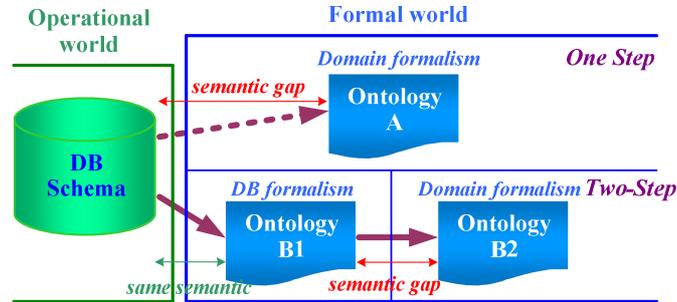

**Fig. 2.** One-step and two-step formalization approaches

For the one-step approach, the above mentioned mappings are carried out in one step. In its mapping file, it uses domain ontology as the target ontology that the database schema maps to. Such an approach is sufficient when the domain formalism is close to the database schema; however, it has certain defects in bridging a large semantic gap. E.g. formalizing the human:weighs in Fig. 1 as a blank node in one step requires to state the complex joins as embedded SQL statements in the RDB-to-RDF mapping file [18], which implies the mapping from RDB semantic to domain formalism is not explicitly stated and carried out in a non-semantic way.

Another issue exists in the one-step mapping is that bringing the semantic mapping in the RDB-to-RDF mapping process may potentially jeopardize its performance. Translating SPARQL queries in the RDF domain into efficient SQL queries in the RDB domain is not an easy task. In [16], it is reported that existing RDB-to-RDF translation tools are not performing well when running on a large database of a real world application. We also discovered similar problems, i.e. processing a SPARQL query with many OPTIONAL statements is very slow with most existing tools. In formalizing the "CLL Form" table in Fig. 1, suppose this table contains many properties which need to be formalized as blank nodes associated to the "Natperson" instance (it is the case in our application), then each property needs to provide its own SQL statement to describe the joins to the "Natperson". This will result in a lengthy and implicit mapping file; meanwhile also bring in many join statements in the generated SQL script. When those properties are queried in the optional blocks of SPARQL queries, it is almost impossible to generate efficient SQL queries. Taking these concerns into account, we deem the one-step formalization may only apply to domain formalisms which are close to the database schema.

The two-step approach separates the RDB-to-RDF mapping from the formal semantic mapping. The first step is about creating a database formalism by mapping an RDB schema to a Data Definition Ontology (DDO) (Ontology B1). Data stored in RDB is formalized with terms from DDO. The DDO is generated based on the RDB schema and they share the same semantics. The second step is an RDF-to-RDF mapping where DDO (Ontology B1) semantics are further converted to Domain Ontology (DO) (Ontology B2) semantics. We apply N3 rules [23] converting the data represented with implicit DDO semantics to those represented with more expressive and explicit DO semantics. The N3 rules allow usage of many complex operations, so that the DO Ontology B2 can be more expressive than the DO Ontology A in the one-step approach, and have clean semantics detached from original data sources. This conversion process can also generate proofs (by the reasoning engine) for validation by independent proof checkers to build trust.

Having stable DDO ontologies to represent the semantics of a data source makes the formalizations in the two-step approach more reusable than those from the one-step approach. The two-step approach allows the data source to be easily integrated in other domains by reusing the DDOs with additional DDO to DO conversion rules. While the one step approach has to create new mapping files to state complex mappings between the RDB schemas and the new DO formalisms, which is more complex and less explicit.

## 4 Structure of Two-Step Formalization

### 4.1 Overview of Two-step Formalization

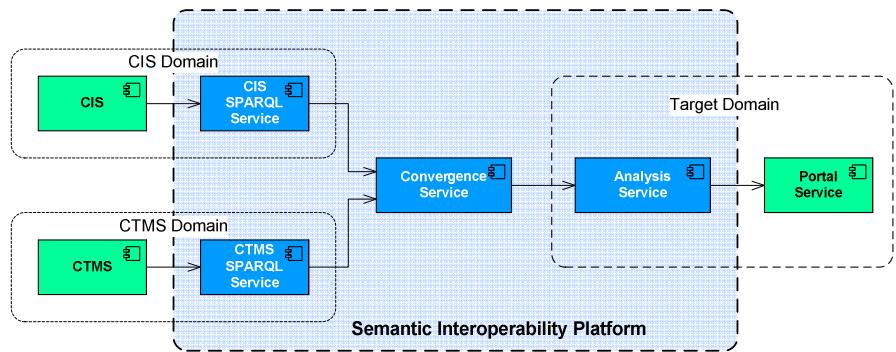

**Fig. 3.** Structure of two-step formalizing approach

Figure 3 shows the structure of the two-step formalization approach in the clinical domain. CISs, as well as other systems, e.g. a clinical trial management system (CTMS) in this figure, can be formalized and integrated together in this platform. The direction of the arrows indicates the data flow in the formalization process.

In the first step, the system formalizes the CIS and CTMS with their local ontology respectively. SPARQL endpoints are built on top of each system, whereupon SPARQL queries can be executed. In the second step, the returned results from the first step are converted to results represented in the ontology of the target domain by the convergence service. The conversion process of the two systems is carried out separately but the converted results are aggregated. Next, analysis rules are applied on the integrated data to deduce further knowledge, and the final results are queried by a portal service to provide decision support for physicians.

### 4.2 Formal Languages

The World Wide Web Consortium (W3C) establishes a set of standards which allow to formally describe, query and exchange information with explicit semantics. The

Semantic Web Stack [24] describes how to use these technologies to build up the Semantic Web where data from different sources are semantically interoperable. Our formalization is based on the formal languages listed in the semantic web stack.

We use URIs to uniquely identify resources, e.g. a patient ID is processed as a URI (see later in Table 2), so that different resources can be distinguished and integrated by URIs. We use RDF triples to represent formalized data and use SPARQL queries to restrict, project and retrieve RDF triples. We apply RDF, RDFS, and OWL ontologies to build our own ontologies, and existing ontologies are also reused.

The ontologies and the RDF triples used in our approach are represented in N3 [19] syntax. N3 is compact and much more human readable compared to RDF/XML [20], and it can be used to express the entire Semantic Web stack formalisms, from RDF triples over rules to proofs, whereas RDF/XML can only be used to express data and ontologies. The conversion and analysis processes (see Fig. 3) are both also expressed as N3 rules [23] and executed by Euler YAP Engine (EYE) [21], an open source reasoning engine. The EYE engine can also generate proofs to build trust.

### 4.3 First Formalization Step

The first formalization step targets on extending operational systems with a semantic interface to expose their data in a formalized format. It could either statically dump the entire content of a system into an RDF store, or build a SPARQL endpoint to dynamically translate SPARQL queries to SQL queries.

A local ontology needs to be developed in both of the above mentioned solutions. As stated in the previous sections, the local ontology in our two-step approach is developed based on the local database schema and we name it as Data Definition Ontology (DDO). We generate DDOs from the local database schemas with a one-to-one mapping following the rules below:
- a database table is mapped to an RDFS class (rdfs:Class);
- a database table column is mapped to an RDF property (rdf:Property);
- the database data type of a field is mapped to the XSD data type range class of the property, one exception is that if a field is a foreign key, its range is the class that the foreign key points to;

**Table 1.** Sample Data Definition Ontology

```
@prefix hosstay: <http://…HospitalStay#>.
…
  hosstay:HospitalStay a rdfs:Class.
  hosstay:persnr a rdf:Property;
        rdfs:domain hosstay:HospitalStay;
        rdfs:range patient:Patient.
  hosstay:hasCLLForm a rdf:Property;
        rdfs:domain hosstay:HospitalStay;
        rdfs:range cllform:CLLForm.

  patient:Patient a rdfs:Class.
  patient:persnr a rdf:Property;
        rdfs:domain patient:Patient;
        rdfs:range natperson:Natperson.

  natperson:Natperson a rdfs:Class.

  natperson:persnr a rdf:Property;
        rdfs:domain natperson:Natperson;
        rdfs:range person:Person.
  natperson:name a rdf:Property;
        rdfs:domain natperson:Natperson;
        rdfs:range xsd:Literal.

  cllform:CLLForm a rdfs:Class.
  cllform:weight a rdf:Property;
        rdfs:domain cllform:CLLForm;
        rdfs:range xsd:long.

  cllform:date a rdf:Property;
        rdfs:domain cllform:CLLForm;
        rdfs:range xsd:Date.
```

**Table 2.** Data Definition Ontology SPARQL Query and Sample Results

| SPARQL Query: | Sample results |
|---|---|
| CONSTRUCT<br>{?hosstay hosstay:hasCLLForm ?cllForm.<br>  ?hosstay hosstay:persnr ?patient.<br>  ?cllForm cllform:weight ?weight.<br>  ?cllForm cllform:height ?height.<br>  ?cllForm cllform:date ?date.<br>  ?patient patient:persnr ?person.<br>  ?person natperson:name ?name.<br>}<br>WHERE<br>{?hosstay hosstay:hasCLLForm ?cllForm.<br>  ?hosstay hosstay:persnr ?patient.<br>  ?cllForm cllform:weight ?weight.<br>  ?cllForm cllform:height ?height.<br>  ?cllForm cllform:date ?date.<br>  ?patient patient:persnr ?person.<br>  ?person natperson:name ?name.} | <http://…HospitalStay/1553541#this><br>  hosstay:hasCLLForm<br>  <http://…CLLForm/359685#this>.<br><br><http://…HospitalStay/1553541#this><br>  hosstay:persnr <http://…Patient/644007#this>.<br><br><http://…CLLForm/359685#this><br>  cllform:weight "72"^^xsd:long.<br><br><http://…CLLForm/359685#this><br>  cllform:height "170"^^xsd:long.<br><br><http://…CLLForm/359685#this><br>  cllform:date "2012-01-01"^^xsd:date.<br><br><http://…Patient/644007#this><br>  patient:persnr <http://…Natperson/644007#this>.<br><br><http://…Natperson/644007#this><br>  natperson:name "Anonymize"^^xsd:Literal. |

Table 1 shows part of a DDO in N3 format, which is generated from the tables in Fig. 1. Note that the prefix declaration of the displayed DDO is incomplete, since the resources are not public.

We build up a SPARQL endpoint on top of the RDB in the ORBIS® CIS system. SPARQL queries can be executed on the SPARQL endpoint using the DDO. Table 2 shows an example of using a SPARQL query to look up data stored in the tables displayed in Fig. 1. The SPARQL query uses the DDO defined in Table 1. For performance concerns, we also opted to use a set of simplified DDO queries instead of complex, monolithic ones. By staying close to the database schema and taking the aforementioned policies, the DDO queries can easily be translated into efficient SQL queries. Sample results (after anonymization) are also shown in Table 2. These results will be processed by conversion rules so as to be expressed with Domain Ontology.

### 4.4 Second Formalization Step

**Domain Ontology (DO)**

The second formalization step targets on translating the data formalized based on the DDO into data formalized with DO semantics. Therefore, building up the DO is the prerequisite of this step.

The DO is created independent of a database schema, but based on domain knowledge to be captured, e.g. a term list created by a clinical domain specialist. When we need a new ontology to cover a new domain, we first search for existing resources. We use Dublin Core Element Set (http://purl.org/dc/elements/1.1/) to declare an ontology provenance header in all namespaces. SKOS (http://www.w3.org/2004/02/skos/core#) is used for terminology/code mapping, e.g. SNOMED CT and ICD10, and for linking clinical codes to classes. FOAF (http://xmlns.com/foaf/0.1/) and Contact (http://www.w3.org/2000/10/ swap/pim/contact#) are used for describing administrative data. We also use some elements of the NASA SWEET ontologies series (http://

sweet.jpl.nasa.gov/2.0/), to describe e.g. quantities and units. Up till now, we created a series of 115 ontologies, all of these ontologies are open source and publicly accessible. These ontologies are divided into a non-clinical series[1] and a clinical series[2].

Having both DDO and DO formalisms, we can now create conversion rules to map data expressions from DDO semantics to DO semantics.

**Conversion Rules**

Taking the example in Fig. 1, the N3 rules in Table 3a have been created to convert DDO triples in Table 2 into DO triples in Table 3b. '=>' stands for log:implies [22], its subject (the left side graph of '=>') is the antecedent graph, and the object (the right side graph) is the consequent graph. N3 rules allow for using blank nodes in the conclusion graph, thus allowing the creation of new anonymous instances. The conversion rules, as well as the analysis rules introduced in later sections, are expressed as N3 rules and executed by the EYE reasoning engine [21]. A rule can pick up the conclusion of another rule, and a set of functions can be used (see Section 5.1).

**Table 3a.** Sample Conversion Rules

| Rule 1: | Rule 3: |
|---|---|
| {?person a natperson:Natperson} => {?person a human:Person}. | {?hosstay hosstay:hasCLLForm ?cllForm. ?hosstay hosstay:persnr ?patient. ?cllForm cllform:weight ?weight. ?cllForm cllform:date ?date. ?patient patient:persnr ?person.} => {?person human:weighs [ quant:hasValue ?weight; event:hasDateTime ?date; quant:hasUnit units:kilogram]}. |
| **Rule 2:** | |
| {?person natperson:name ?name} => {?person foaf:familyName ?name}. | |

**Table 3b.** Sample Conversion Results

<http://…Natperson/644007#this> a human:Person.
<http://…Natperson/644007#this> foaf:familyName "Anonymize"^^xsd:Literal.
<http://…Natperson/644007#this> human:weighs [ quant:hasValue "72"^^xsd:long; event:hasDateTime "2012-01-01"^^xsd:date; quant:hasUnit units:kilogram].

By executing the conversion rules in Table 3a on the sample results displayed in Table 2, the converted results shown in Table 3b are generated. The Rule 3 generates a blank node for DO predicate human:weighs. Some additional DO properties are used to formalize the weight value, date, and weight unit. A similar rule can be easily developed to generate a blank node for human:hasLength using the same DDO graph in Table 2. As we claimed, with this approach, the DDO SPARQL query is very simple (and easy to be translated into concise SQL); the conversion from DDO to DO is also simple and most importantly in an explicit and semantic way. After the conversion process, data stored in a CIS is represented as RDF graph/triples using the DO. It can be integrated with RDF graphs from other sources to form a bigger graph and apply analysis rules to deduce more knowledge for clinical decision support.

---

[1] http://eulersharp.sourceforge.net/
[2] http://www.agfa.com/w3c/2009/

## 5   Implementations

In this section, we first present a detailed example of using the two-step approach to formalize the calculation of body mass index (BMI). Then we briefly introduce the applications of the two-step approach in two projects.

### 5.1   Formalization of Body Mass Index Calculation

After mapping the clinical data with DO, N3 rules are created to use all kinds of clinical formal data to perform a variety of calculations and other clinical analyses. Examples include the calculation of BMI and evaluation of weight status. Complex tasks such as establishing staging, e.g. calculating Binet Stage for Chronic Lymphocyte Leukemia (CLL) depending on blood measurement values are also implemented in our projects. In addition, Clinical Decision Support for therapies can be performed depending on e.g. genome findings in the case of CLL or the measurement of blood neutrophil granulocytes. This section will take the formalization of BMI calculation as an example to demonstrate how we formalize, integrate data, and perform analysis rules in a formal way in the clinical domain.

**Theorem 1: BMI = human body weight (kg) / (human body length (m))** $^2$

The basic theorem of BMI calculation is rather simple as shown in Theorem 1. However, once it has to be performed in the clinical domain, such a trivial task becomes complex. In order to carry out the mathematical calculation in Theorem 1, many conditions have to be explicitly met, e.g. the calculation only applies to adults, so that the measured person has to be an adult; the measurements of weight and length should be carried out not too far apart in time, etc. The formalization of BMI calculation shown in the rest of this section takes all these constraints into account.

Table 4 shows the analysis rule to formalize the calculation of BMI. The left part of the table shows the analysis rule, and the right part gives some explanation. The first ten lines of the analysis rule is the preparation step, which gets the data to be analyzed: the weight and the height as well as their measurement date, together with the birth date of a patient. Figure 4 uses RDF graphs to illustrate how the required data is collected and processed. After getting the needed data, Step 1 checks whether the weight measurement and height measurement were carried out in a certain time frame. The condition we specified in this paper is stated below; nevertheless, it can also be replaced by other time frames.

$$-7\,days \leq weight\ date - length\ date \leq 2\,years \qquad - \text{Step 1}$$

Step 2 takes the latest date between the weight and length measurement date, and Step 3 calculates the age of the patient during the latest test, there is a separate rule implementing the function below:

$$age = latest\ test\ date - birth\ date \qquad - \text{Step 3}$$

**Table 4.** An Analysis Rule with Explanations

```
{?adult a human:BiologicalAdult;
       organism:hasBirthDateTime ?birthDate.
 ?adult human:weighs [
       quant:hasUnit units:kilogram;                PREPARATION STEP: GETTING DATA
       quant:hasValue ?weightValue;
       event:hasDateTime ?weightDateTime].
 ?adult human:hasLength [
       quant:hasUnit units:meter;
       quant:hasValue ?lengthValue;
       event:hasDateTime ?lengthDateTime].

 (?weightDateTime ?lengthDateTime)
       math:difference ?dif.                        STEP 1: CHECKING DELAY BETWEEN WEIGHT
 ?dif math:notGreaterThan "P2Y"^^xsd:duration;      AND LENGTH MEASUREMENTS
      math:notLessThan "-P7D"^^xsd:duration.

 (?weightDateTime ?lengthDateTime)                  STEP 2: CALCULATING LATEST DATE TIME
       e:max ?maxDateTimeNumeral.                   OF AN EXAMINATION
 (?lexical xsd:dateTime)
       log:dtlit ?maxDateTimeNumeral, ?dateTime.

 (?adult ?dateTime) time:calculatingAge ?ageInYears. STEP 3: CALCULATING AGE AT MOMENT OF
                                                    LATEST EXAMINATION

 ?ageInYears math:notLessThan 18.                   STEP 4: CHECK ADULT AGE

 (?weightValue (?lengthValue 2) !math:exponentiation) STEP 5: CALCULATION OF ADULT BODY MASS
       math:quotient ?bodyMassIndexValue}           INDEX BASED ON THEOREM 1
 =>
 {?adult human:hasBodyMassIndex [                   FINAL STEP: GETTING RESULTS
     quant:hasUnit units:kilogramPerMeterSquare;
     quant:hasValue ?bodyMassIndexValue;
     event:hasDateTime ?dateTime]}.
```

Step 4 checks whether the patient could be considered as an adult (arbitrarily age>18), and Step 5 executes the equation displayed in Theorem 1 to get the BMI. The final results are generated below the '=>' sign.

Figure 4 shows the conversion, integration and analysis of RDF graphs for the BMI calculation. In the RDF graph, each triple {subject, predicate, object} is represented as a node-arc-node link. For example, the first arc in the DDO graphs represents a triple {?hospitalCaring hosstay:hasCLLForm ?cllForm.}.

The dataflow starts from four DDO graphs which are results of DDO SPARQL queries and similar to those explained in Table 2. These graphs are converted to DO graphs by executing conversion rules, similar to the examples shown in Table 3. The four graphs in DO terms are integrated together providing they contain the same URI for the ?person variable. Analysis rules are applied on the integrated graph as introduced in Table 4 and the generated BMI graph is shown in the dotted square. The dataflow presented in Fig. 4 is coordinated by the semantic interoperability platform shown in Fig. 3.

In practice, we also have rules to convert different units in height measurement; it is not introduced in this example due to page limitations. In this BMI calculation process, we may observe the huge semantic gap between the operational world and the domain ontology in the clinical domain. A seemingly simple example turns out to be rather complex when taking the clinical correctness explicitly into account. Semantically processing the data with correct clinical knowledge removes the ambiguity of the data and guarantees the correctness of the data. The semantically processed data can further be parsed and processed by a computer automatically.

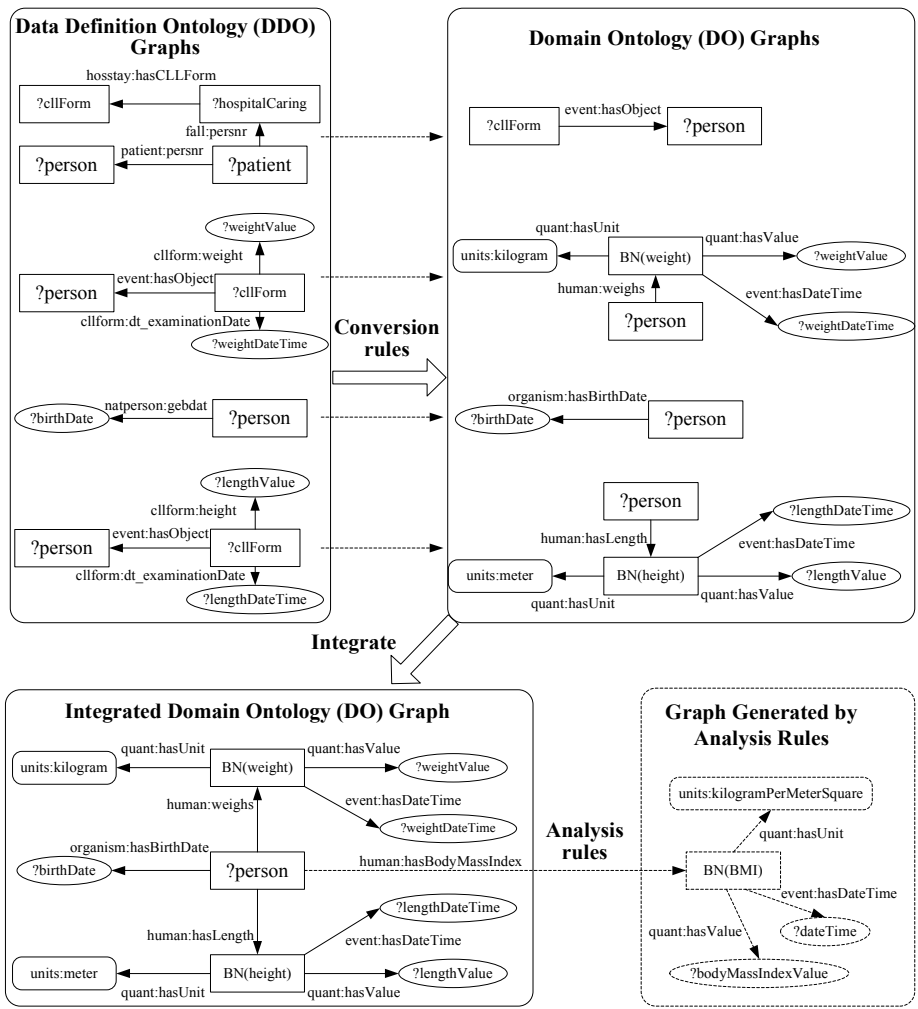

**Fig. 4.** Data Flow of Body Mass Index Formalization – An RDF Graph View

### 5.2 Applications

The two-step formalization approach presented in this paper has been implemented and validated in two projects we have coordinated. The EU's DebugIT project [9] investigates semantic formalization and integration of lab data from seven hospitals located in different countries across Europe to monitor antibiotic resistance. SPARQL endpoints are built on top of each clinical site respectively; each SPARQL endpoint generates its own DDO based on the data structure of their local clinical information system. The DO is developed and mappings from DDO to DO are also created at each site. After integrating and analyzing the data, many clinical questions can be an-swered, e.g. bacterial resistance to a particular antibiotic drug in a particular case during a particular period is reflected in the DebugIT Web Portal, where both the resistance level of each site and the averaged level over the seven sites are stated.

In the HIT4CLL project [10], we aim to provide clinical support and trial selection for Chronic Lymphocyte Leukemia (CLL). We formalized clinical data stored in ORBIS® CIS system and integrated it with a clinical trial management system. The integration is achieved using URIs for clinical trials. Table 5 summarizes the formalizations we carried out in this project: DDOs are generated for more than 20 relational tables, which contain 508 columns in total. These tables store more than 1.3 billion rows of data, roughly equivalent to 33 billion (1.3 billion×508/20) triples when every table field in the database is considered as a triple. The data we process for each patient starts with retrieving DDO graphs from this data source. The retrieved DDO graphs for each patient contain on average 14,000 triples, and in the end generating an EHR graph (the final DO graph after applying all the analysis rules together with the original DDO graph) with around 32,000 triples. In this project we applied 47 Domain Ontologies of which 41 are developed internally and 6 are external ones. These 47 DOs contains 15,770 triples. About 100 conversion rules and 300 analysis rules are created (e.g. the BMI calculation shown in Table 4 is considered as one analysis rule).

**Table 5.** Formalization Statistics

| Sources | Numbers |
|---|---|
| Data Source | |
| Number of formalized tables | 20 |
| Number of formalized columns | 508 |
| Number of rows | 1.3 billion |
| Equivalent in triples | 33 billion |
| Patient Graph | |
| Size of DDO graph (on average) | 14,000 triples |
| Size of EHR graph (on average) | 32,000 triples |
| Ontologies | |
| Number of Domain Ontologies | 47 |
| Internal | 41 |
| External | 6 |
| Size of Domain Ontologies | 15,770 triples |
| Rules | |
| Number of Conversion Rules | ~100 |
| Number of Analysis Rules | ~300 |

The whole process, from selecting data from RDB to generating an EHR graph takes 12.5 seconds on average on a standard server. The final results (the EHR graph) are queried and displayed by the HIT4CLL web portal in order to provide a better interface for clinicians. Information contained in EHR graphs is displayed in the portal in sorted sections, such as demographics, diagnosis, etc. (see Fig. 5). Both calculated results, like the BMI, or original data, like different lab results are displayed.

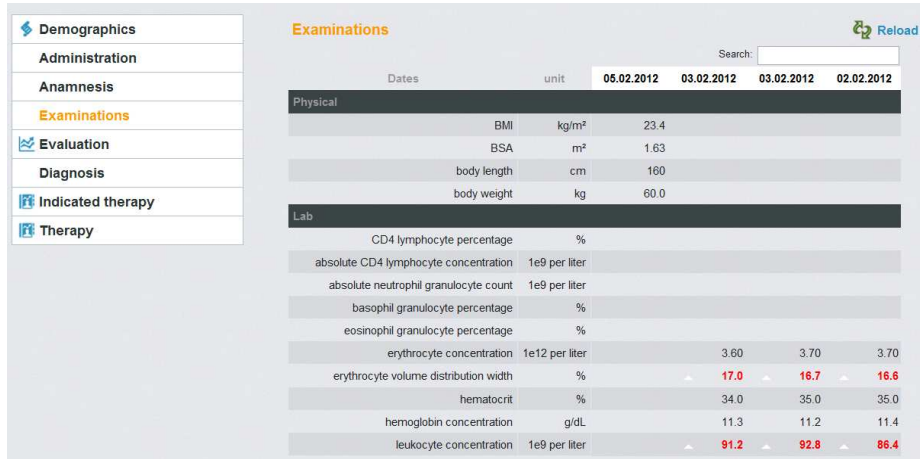

**Fig. 5.** HIT4CLL Web Portal

The EHR graphs for each patient can also be aggregated together to support population query: a user may query a group of patients to retrieve population views on the calculated results, e.g. the distribution of BMI (calculated by analysis rules) in a population group with certain age. In order to support such queries, the graphs of the target population group need to be aggregated together first (Aggregation), and then rules can be applied on the aggregated graph to generate population views (Calculation). Such rules are developed in the HIT4CLL project to provide a set of population views. The scalability test on different size of population group is carried out and the results are displayed in Fig. 6. The slices of aggregation and calculation are aforementioned, they both requires time to retrieve data (Retrieve data) and time (Reasoning) to resolve the data or reason the rules; those are displayed in separated bars in Fig. 6. The time spent on generating EHR graph for each patient (around 12.5s for one patient) is not included in this test, these graphs are afore generated and cached for the aggregation. The performance test shows that most of the time is spent in aggregating the graphs, mainly consumed in retrieving the data. It also shows that the system exhibits a liner performance and consumes less than six minutes to generate a set of population views for a group sized with 1280 patients.

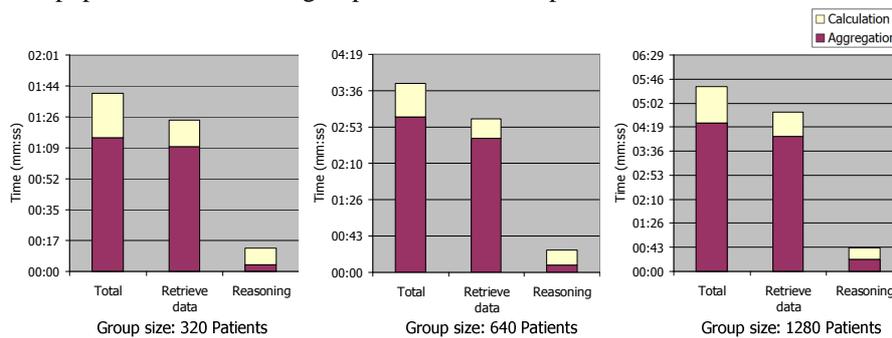

**Fig. 6.** Scalability Test in Retrieving Population View

Through the above mentioned two projects, we have implemented our two-step formalization approach and tested its performance and scalability. The drawback of the two-step formalization approach is that the conversion process is carried out from DDO to DO in a one way direction. We have investigated query rewriting methods so as to translate queries represented with DO into counterparts represented with DDO. We did not reach our goal due to the fundamental un-scalability in translating abstracted knowledge (DO) to many concrete ones (DDO), still allowing for new systems to join. Therefore in our solution, we use a query generator using predefined DDO SPARQL query templates to retrieve data from the connected data sources. The current solution meets the requirements of the two projects we participated in: using semantic formalization to process data and provide answers to a set of predefined parameterised questions.

# 6   Conclusions

Semantic processing and interoperation of clinical data is becoming a rising request for Healthcare IT systems. While most of the research performs formalization on standard EHRs, we suggest that in order to utilize clinical data in the best possible way, one should formalize data as soon as possible. Therefore, we start our formalization directly from the data structure of clinical information systems (CIS). By this approach, our solution is not relying on the existing EHR standards and can be applied on any CIS.

We observed in the clinical domain, especially in tackling real world clinical cases using lab and diagnosis data stored in a CIS, that there is a large semantic gap between the local database semantics and the domain semantics. Directly mapping a local database schema to the domain ontology has to formalize local data structure and carry out the local-to-domain semantic transformation in a single step, which largely increase the complexity and may degrade the performance.

Therefore we argue that such a one-step formalization approach is not suitable for certain clinical applications. In this paper, we reported our practice in using a two-step formalization approach which separates the tasks of formalizing local data structure and the conversion from the local data definition ontology to the domain ontology. Using simplified Data Definition Ontologies and Data Definition Queries guarantees the efficiency in retrieving data from a CIS. Our formalization is built with the formal languages that construct the Semantic Web Stack. To the best of our knowledge, this is the first practice supporting real-world application processing large scale clinical practice data, using formal languages in the whole data flow and implementing most of the layers of the Semantic Web Stack.

The two-step formalizing approach is successfully implemented and tested in two projects [9] [10]. The structure of this approach is presented in this paper; the usage of rules and the corresponding data flow in this approach is also explained using BMI calculation as an example. For the future work, we will keep validating our approach in other projects to provide more clinical decision supports and further test and improve its performance and scalability to process larger patient groups.

## Acknowledgement

This research has been supported by the EU-IST-FP7 DebugIT project and the HIT4CLL project. The authours would like to explicitly thank Hans Cools for his important contributions to the formalization approach.